# Ultrafast preparation and strong-field ionization of an atomic Bell-like state

Authors:

Sebastian Eckart, Daniel Trabert, Jonas Rist, Angelina Geyer, Lothar Ph. H. Schmidt, Kilian Fehre, Maksim Kunitski

Institut für Kernphysik, Goethe-Universität, Max-von-Laue-Straße 1, 60438 Frankfurt, Germany

Abstract: Molecules are many body systems with a substantial amount of entanglement between their electrons. Is there a way to break the molecular bond of a diatomic molecule and obtain two atoms in their ground state which are still entangled and form a Bell-like state? We present a scheme that allows for the preparation of such entangled atomic states from single oxygen molecules on femtosecond time scales. The two neutral oxygen atoms are entangled in the magnetic quantum number of their valence electrons. In a time-delayed probe step, we employ non-adiabatic tunnel ionization, which is a magnetic quantum number-sensitive mechanism. We then investigate correlations by comparing single and double ionization probabilities of the Bell-like state. The experimental results agree with the predictions for an entangled state.

One-Sentence Summary: We prepare two spatially separated entangled atoms by dissociating $O_2$ and examine tunnel ionization of these twin atoms.

Main Text:

Tunneling *(1–3)* and entanglement *(4–6)* are two of the most intriguing phenomena of quantum mechanics. Tunneling occurs because there is a non-vanishing probability for quantum-mechanical particles to transverse a classically forbidden region, which is referred to as tunnel barrier. In experiments, very strong laser fields can liberate an electron from a single atom or molecule by tunnel ionization *(7)*. For circularly polarized light the tunnel barrier rotates in the polarization plane which leads to intriguing non-adiabatic dynamics in the classically forbidden region *(3)*. This results in a significant dependence of the tunneling probability on the magnetic quantum number $m$ of the electron *(8, 9)*. Thus, non-adiabatic tunneling acts as a polarization filter preferring electrons with $m = -1$ for transmission through the tunnel *(10)*.

Entanglement is based on the correlation of the quantum-mechanical wave function $\Psi$. In real world measurements, only the modulus square of the wave function $|\Psi|^2$ can be accessed directly. In quantum mechanics, correlations can exist not only – as for classical correlations – on the level of $|\Psi|^2$ but on the level of $\Psi$ as well. This has far-reaching consequences, which were questioned by Einstein, calling it "spooky action at a distance" *(11, 12)*. The realization that the quantum realm violates local realism was groundbreaking *(13–15)* and gave rise to technologies that harness entanglement for quantum information protocols *(16)*. Famous examples of entangled states are Bell states, which can be prepared in atoms by resonant optical transitions *(17)*. For resonant transitions the spectral width of the transitions limits the speed of preparation.

We experimentally demonstrate the preparation of a Bell-like state from single oxygen molecules in their ground state using femtosecond laser pulses. The excitation leads to the dissociation of the molecule resulting in two neutral oxygen atoms in their ground state that move in opposite directions with a velocity of about 2500 m/s. The spatially separated atoms are entangled in the magnetic quantum number $m$ of their valence electrons. We use the following notation for the prepared Bell-like state:

$$\Psi^+_{-00-} = \frac{1}{\sqrt{2}}(|m_{-1}\rangle_A|m_0\rangle_B + |m_0\rangle_A|m_{-1}\rangle_B) \quad (1)$$

Here, $|m_{-1}\rangle_A$ indicates that in the atom at site A, there are two electrons with a magnetic quantum number of $m = -1$ in the 2p orbital while there is only one electron with $m = 0$ and one electron with $m = +1$ (see Fig. 1). The other notations are analogous. The quantization axis of the angular momentum after preparation is defined by the former molecular axis, which is experimentally accessible from the velocity vectors of the two neutral oxygen atoms *(18)*.

The experimental scheme for the ultrafast preparation of a Bell-like state from an oxygen molecule is illustrated in Fig. 1. The incident pump pulse excites the $\pi^*_{2p}$ electron with $m = +1$ to the $\sigma^*_{2p}$ level with $m = 0$ (Fig. 1, C). Subsequently, the molecular state $1^3\Pi_u$ dissociates into two oxygen atoms in their ground state *(19, 20)*. After the atoms are spatially separated, they are in the well-defined atomic states $^3P_1$ and $^3P_2$ as exemplarily illustrated in Fig. 1, D *(21)*. Due to the conservation of angular momentum and the helicity of the pump pulse, the sign of $m$ is defined. Thus, the atom in the $^3P_2$-state in Fig. 1 is in the state $|m_{-1}\rangle$. Consequently, the atom in the state $^3P_1$ in Fig. 1 must possess two 2p electrons with $m = 0$ and is thus indicated using the notation $|m_0\rangle$. Our convention regarding the sign of $m$ is chosen such that electrons with $m = -1$ would be counter-rotating in a semi-classical picture with respect to the rotational direction of the electric field of the probe laser pulse *(8, 10, 22)*. The left and the right half of Fig. 1, D together illustrate the entangled atomic state from Eq. 1. Only 100 fs after the pump pulse has triggered dissociation, the two neutral atoms are separated by about 5 Ångstrom, which is far enough to safely neglect classical interaction between them *(23)*.

Since the prepared Bell-like state is entangled in the magnetic quantum number $m$ of its electrons, its quantum properties can only be accessed by an $m$-selective interaction. Strikingly, non-adiabatic tunnel ionization is not only $m$-selective but also an ultrafast process that occurs on attosecond time scales *(24, 25)*. To this end, we use a circularly polarized light field as a probe pulse and thereby exploit the fact that non-adiabatic tunneling *(8, 10)* strongly prefers electrons with $m = -1$. In order to test the quantum-properties of this Bell-like state, we study different orientations of the molecular axis with respect to the light propagation direction and compare the probability for the single ionization of one of the atoms with the probability to singly ionize both atoms of the Bell-like state.

In case the two atoms of the prepared Bell-like state dissociate along the light propagation direction, a correlation in $|\Psi|^2$ cannot be distinguished from entanglement in $\Psi$. This can be explained as follows: if we selectively liberate electrons in the $|m_{-1}\rangle$-state and observe one ionized atom and one neutral atom after the laser pulse has hit the atom, then we would know that the ionized atom must have been in the $|m_{-1}\rangle$-state and that the neutral atom must be in the $|m_0\rangle$-state. Thus, when the molecular axis is aligned along the light propagation direction, a correlation in $|\Psi|^2$ cannot be distinguished from entanglement in $\Psi$.

Classical correlations and entanglement are distinguishable if the oxygen molecule does not dissociate along the light-propagation direction of the probe pulse. For the sake of the argument, we consider a molecule that dissociates along a direction that is tilted by 45° with respect to the light's quantization axis as illustrated in Fig. 2, A. Note that in the probe step, the light propagation direction defines the quantization axis. This new quantization axis is different to the quantization axis of the magnetic quantum number $m$, which is defined by the molecular axis of the former molecule. Consequently, the new quantization axis implies a new set of magnetic quantum numbers which are referred to as $m'$. Thus, tunneling that is driven by the probe pulse is $|m'_{-1}\rangle$-selective. The projection to the new basis can be written using:

$$|m_{-1}\rangle_A = a|m'_{-1}\rangle_A + b|m'_0\rangle_A + c|m'_{+1}\rangle_A \tag{2}$$
$$|m_0\rangle_A = d|m'_{-1}\rangle_A + e|m'_0\rangle_A + f|m'_{+1}\rangle_A \tag{3}$$

The definitions for $|m_{-1}\rangle_B$ and $|m_0\rangle_B$ are analogous. For $\alpha = 45°$ the coefficients are given by $a = \left(\frac{1}{\sqrt{8}} + \frac{1}{2}\right)$, $b = \frac{1}{2}$, $c = \left(\frac{1}{\sqrt{8}} - \frac{1}{2}\right)$, $d = -\frac{1}{2}$, $e = \frac{1}{\sqrt{2}}$ and $f = -\frac{1}{2}$ (see Supplementary Material for details). The state from Eq. 1 can be expressed using the new basis:

$$\Psi^+_{-00-} = \frac{1}{\sqrt{2}}\Big[\big(a|m'_{-1}\rangle_A + b|m'_0\rangle_A + c|m'_{+1}\rangle_A\big)\big(d|m'_{-1}\rangle_B + e|m'_0\rangle_B + f|m'_{+1}\rangle_B\big) + \big(d|m'_{-1}\rangle_A + e|m'_0\rangle_A + f|m'_{+1}\rangle_A\big)\big(a|m'_{-1}\rangle_B + b|m'_0\rangle_B + c|m'_{+1}\rangle_B\big)\Big] \tag{4}$$

Let us do a quantum-mechanical thought experiment and apply the laser only at site A (in contrast to bound molecules, this is possible in our case because the two atoms are spatially separated). Let us further assume that we ionize all parts of the wave function at site A that are in the $|m'_{-1}\rangle_A$-state, and could not ionize any other state, then the single ionization probability would be given by $\frac{1}{2}(a^2 + d^2)$ (see Supplementary Material and Fig. S2 for details). Due to entanglement, ionization at site A would result in a modified wave function for the singly ionized Bell-like state as illustrated in Fig. 2, B. In a next step we could use another laser pulse that is applied at site B which can lead to the subsequent ionization of the remaining atom at site B.

With a certain probability, we would have created two singly charged oxygen ions (see Fig. 2, C). The quantum-mechanical prediction for the probability for double ionization by liberating one electron with $|m'_{-1}\rangle$ at site A and one electron with $|m'_{-1}\rangle$ at site B is given by $\frac{1}{2}(ad + da)^2 = 2a^2d^2$ (see Table S1 for details). This result, that is based on Eq. 1, is in contrast to the expectation of any classical theory based on local realism. For comparison, we look at a classical state, which is described by $\Psi^+_{-0} = |m_{-1}\rangle_A|m_0\rangle_B$ in 50% of the cases and in the other 50% it is modeled by $\Psi^+_{0-} = |m_0\rangle_A|m_{-1}\rangle_B$. If we also assume here that we liberate all electrons which are in the $|m'_{-1}\rangle$-state at site A, this results in an ionization probability of $\frac{1}{2}a^2 + \frac{1}{2}d^2$, which is identical to the quantum-mechanical expectation. However, the probability of ionizing both atoms is $\frac{1}{2}(ad)^2 + \frac{1}{2}(da)^2 = a^2d^2$, which differs from the quantum-mechanical expectation.

Therefore, it is expected, that the Bell-like state from Eq. 1 leads to a single ionization probability that is accurately predicted by the quantum-mechanical as well as the classical model, but the violation of local realism of the prepared state results in a double ionization probability that cannot be explained by classical correlations of the two atoms. The quantum-mechanical description as an entangled pair of atoms captures the correlations of the wave function's amplitudes and reveals that single ionization of one of the atoms alters the wave function of the other atom (Fig. 2, B, C) which manifests in a modified double-ionization probability.

To investigate the Bell-like state from Eq. 1 experimentally, 1.5 picoseconds after the preparation by the pump pulse, we irradiate the prepared state with a circularly polarized femtosecond probe pulse. At this time, the two entangled oxygen atoms are both in a $^3P$ state and have a distance of about 75 Ångstrom. We measure the probability for single ionization of only one as well as both oxygen atoms as a function of the angle $\alpha$, which is the angle between the initial molecular axis and the polarization plane of the probe pulse.

Fig. 3, A shows the experimental results for the single ionization of the Bell-like state from Eq. 1 as a function of $\alpha$. Additionally, we also show the results for the corresponding state that is produced by flipping the helicity of the pump pulse which prepares the energetically degenerate Bell-like state $\Psi^+_{+00+} = \frac{1}{\sqrt{2}}(|m_{+1}\rangle_A|m_0\rangle_B + |m_0\rangle_A|m_{+1}\rangle_B)$. The experimental results are reproduced by our quantum-mechanical model as well as by our classical model (Fig. 3, B). Both models take $m'$-selective tunneling, the absolute ionization probability, the population of the two different Bell-like states as a function of the molecular orientation and the angle dependent dissociation probability into account. Both models have four free parameters which are optimized independently for each model.

Fig. 3, C shows the experimental result for the ionization of both entangled oxygen atoms. The double ionization of the state $\Psi^+_{-00-}$ shows a clear double-hump structure which is not visible for the ionization of the state $\Psi^+_{+00+}$. This is in agreement with our quantum-mechanical modeling that takes quantum entanglement and interference into account (solid lines in Fig. 3, D). For comparison, the dotted lines in Fig. 3, D show the result from our classical model, that includes classical correlations only. The classical model exhibits a less pronounced double-hump structure, which disagrees with the experimental finding. It should be noted that all free parameters of the two models are determined using solely the experimental results for the single ionization of the Bell-like state (see Fig. 3, A and Supplementary Material).

In conclusion, our experimental results reveal properties that are different to the expectations for a classical correlated pair of atoms and agree with the expectations for an entangled, Bell-like state. This supports the perspective that local realism is, as expected, violated in strong field ionization, paves the way towards time resolved studies of entangled states, and highlights the importance of entanglement in chemical systems *(26, 27)* as well as in multielectron processes on attosecond time scales *(6, 28)*.

Acknowledgments:
We thank for fruitful discussions with Reinhard Dörner, Till Jahnke, Markus Schöffler, and Horst Schmidt-Böcking.

Funding:
German Research Foundation (DFG), SPP 1840 QUTIF, Project No. DO 604/29-1 (SE)
German Research Foundation (DFG), SFB 1319 ELCH, Project No. 328961117 (JR, KF)

Author contributions:
S.E designed the experiment. All authors prepared and performed the experiment. S.E. performed the data analysis. S.E., D.T. and J.R. developed the theoretic model. S.E. and K.F. created the figures. All authors contributed to the manuscript.

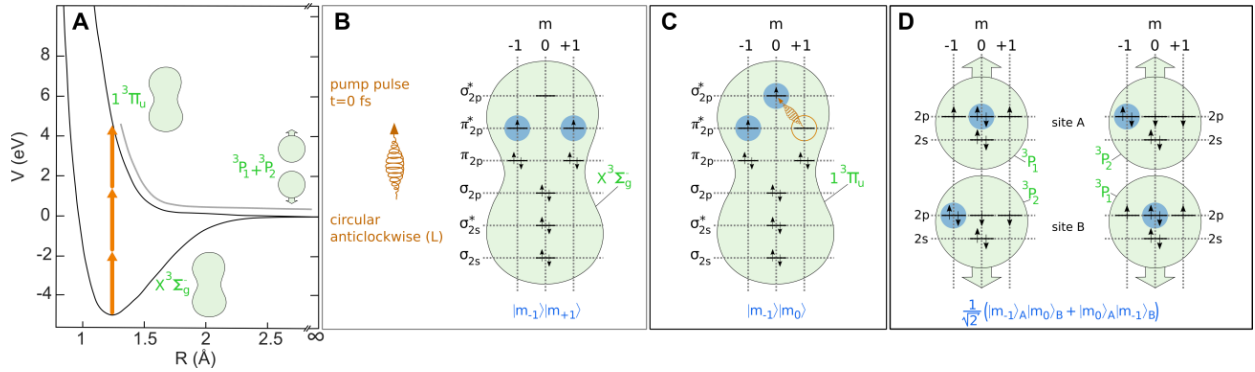

**Fig. 1. Ultrafast preparation of an atomic Bell-like state.** (**A**) An oxygen molecule in its ground state $X^3\Sigma_g$ is excited by three-photon absorption from a circularly polarized pump pulse with anticlockwise-rotating electric field (indicated by "(L)") at a central wavelength of 390 nm and an intensity of $1.0\times10^{14}$ W/cm$^2$. This triggers dissociation of the oxygen molecule via the $1^3\Pi_u$ state and leads to the production of two neutral oxygen atoms. (**B**) Before the three-photon absorption from the pump pulse with an anticlockwise rotating electric field vector, the oxygen molecule in the $X^3\Sigma_g$ state contains one electron with $m=-1$ and one electron with $m=+1$ in the $\pi^*_{2p}$ orbital. (**C**) The pump pulse excites the electron from $\pi^*_{2p}$ with $m=+1$ to $\sigma^*_{2p}$ with $m=0$. This leads to the $1^3\Pi_u$ state that contains an excess electron with a magnetic quantum number of $m=-1$. (**D**) Upon dissociation, this produces two oxygen atoms in their ground state, $^3P_1$ and $^3P_2$ which have different, but defined magnetic quantum numbers (see blue filled circles in B-D). Since it is undecided which state is at which site, this gives rise to entanglement in the magnetic quantum number. The Bell-like state can be written as $\Psi^+_{-00-} = \frac{1}{\sqrt{2}}(|m_{-1}\rangle_A|m_0\rangle_B + |m_0\rangle_A|m_{-1}\rangle_B)$.

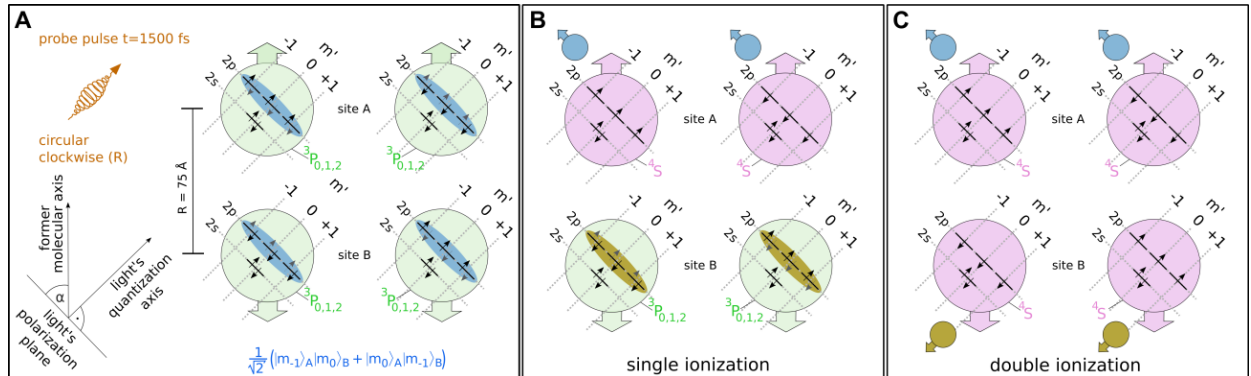

**Fig. 2. Projection of the prepared Bell-like state to a new basis and detection.** (**A**) Molecules that dissociate at an angle of 45° with respect to the polarization plane during the pump step are irradiated with a circularly polarized probe pulse with a clockwise-rotating electric field (indicated by "(R)") that has an intensity of $4.5\times10^{14}$ W/cm$^2$. The probe pulse projects the magnetic quantum number $m$ in the molecular frame onto a new basis $m'$, as illustrated by the gray arrows within the blue shaded area. (**B**) At site A, non-adiabatic tunnel ionization occurs, which strongly prefers $m'=-1$ and acts similarly to a polarizer that projects the wave function on its eigenstates that are defined by the new quantization axis. Single ionization at site A instantaneously affects the wave function at site B (ocher colored area). (**C**) With a certain probability a second electron with $m'=-1$ is liberated at site B by a sequential tunneling process such that both oxygen atoms are singly ionized (double ionization of the Bell-like state).

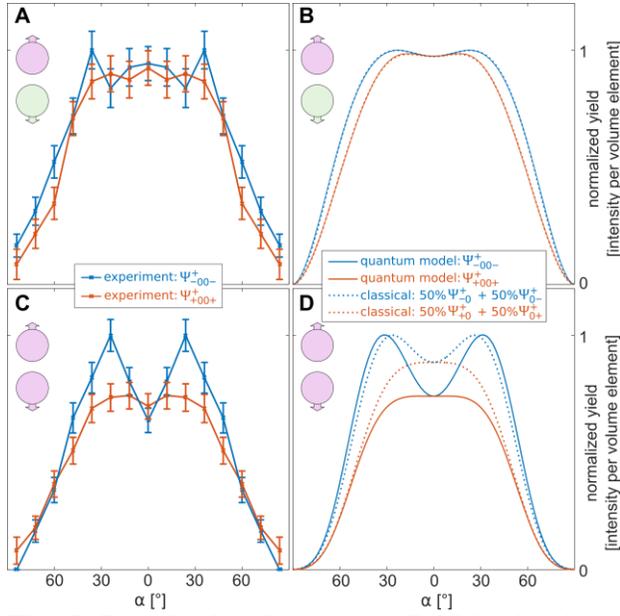

**Fig. 3. Results for the strong-field ionization of two different Bell-like states. (A)** Experimental result for the ionization of one of the two atoms. The blue curve shows the ionization probability as a function of $\alpha$ after preparing the Bell-like state $\Psi^+_{-00-} = \frac{1}{\sqrt{2}}(|m_{-1}\rangle_A|m_0\rangle_B + |m_0\rangle_A|m_{-1}\rangle_B)$. The red curve shows the same for the Bell-like state $\Psi^+_{+00+} = \frac{1}{\sqrt{2}}(|m_{+1}\rangle_A|m_0\rangle_B + |m_0\rangle_A|m_{+1}\rangle_B)$. **(B)** The prediction of our quantum-mechanical model (solid line). The dotted lines show the result using our classical model, which uses a classical state that is described by $\Psi^+_{-0} = |m_{-1}\rangle_A|m_0\rangle_B$ in 50% of the cases and otherwise by $\Psi^+_{0-} = |m_0\rangle_A|m_{-1}\rangle_B$ (the notations for $\Psi^+_{+0}$ and $\Psi^+_{0+}$ are analogous). **(C)** and **(D)** The same as A and B but for the single ionization of both of the two atoms. The quantum-mechanical model (solid line) and the classical model (dotted line) lead to different predictions. The quantum-mechanical model shows better agreement with the experimental data. The experimental data has been symmetrized. Intensity per volume element indicates that the measured yield has been divided by $\cos(\alpha)$. Error bars show the standard deviation of the statistical error only.

## Supplementary Materials

**Materials and Methods**

Laser setup
The optical setup is based on a laser that operates at a central wavelength of 780 nm (KMLabs Dragon, 40-fs FWHM, 8 kHz). The pulses that are used as pump pulses are frequency doubled in a 200 µm $\beta$-barium borate crystal producing laser pulses at a central wavelength of 390 nm. The probe pulses have a central wavelength of 780 nm. The intensity, the ellipticity and the main axis of the polarization ellipse of both pulses can be adjusted independently for the pump and the probe pulses. Both pulses were circularly polarized. The pump-probe delay of 1.5 picoseconds was set using a micrometer delay stage. Both laser pulses were focused by a spherical mirror (f=80 mm) onto a cold supersonic jet of molecular oxygen which was created by expanding oxygen gas through a 30 µm diameter nozzle into vacuum. The helicity of the circularly polarized pump pulse was inverted every 2 minutes to minimize systematic errors. In Fig. 1 the case of a pump pulse with anticlockwise-rotating electric field is illustrated (indicated by "(L)"). Throughout the experiment, the helicity of the probe pulses was not changed and a circularly polarized pulse with clockwise-rotating electric field was used (indicated by "(R)"). The optical setup is the same as in Ref. *(10)*. The intensity of the pump pulse of $1.0\times10^{14}$ W/cm$^2$ was calibrated from the shift of the above threshold ionization peaks as a function of the laser intensity, which is due to the change in ponderomotive energy. The intensity of the probe pulses at a central wavelength of 780 nm was obtained from the measured drift momentum of the electron and found to be $4.5\times10^{14}$ W/cm$^2$. The uncertainty of the absolute intensity for the pump and the probe pulse is estimated to be 20%.

Particle detection
We use a COLTRIMS reaction microscope *(29)* to detect up to two singly charged oxygen ions in coincidence with one electron. The electron and ion arm of the spectrometer have a length of 378 mm and 67.8 mm, respectively. The charged fragments are guided by a homogeneous electric field (17.3 V cm$^{-1}$) and a homogeneous magnetic field (10.1 G) towards time- and position-sensitive detectors. Each detector consists of a stack of two multi-channel plates (MCPs). The MCPs of the electron and the ion detector have a diameter of 120 mm and 80 mm, respectively. For both detectors, the MCP stack is followed by a three-layer hexagonal delay-line anode (HEX) *(30)*. We found that for $p_y<0$ a.u. the electron detection efficiency was affected by a local inefficiency of the MCP. To minimize systematic errors due to detector inefficiencies, we excluded events with an electron momentum $p_y<0$ a.u. (the gas jet propagates along the $p_y$-direction). The molecular axis is accessible by making use of the axial recoil approximation, which holds in our case because the dissociation is fast compared to the rotation of the $1^3\Pi_u$ state. Therefore, we can infer the former molecular axis by using that the momentum of the detected ion points along the molecular axis at the time the molecule was hit by the pump pulse *(18)*.

Isolation of the neutral dissociation channel in the coincidence pump-probe experiment
The pump pulses lead to dissociation of oxygen molecules without ionizing them. This is evident from Fig. S1 by using that neutral oxygen atoms can still be ionized by the probe pulse *(31)*. For Fig. S1A only the pump pulses were used and the kinetic energy release of the atoms (KER) is shown versus the kinetic energy of the coincidently detected electron. The KER is calculated to be two times the kinetic energy of the first detected oxygen ion.

In strong field ionization, the momentum of the liberated electron can be approximated to be determined by the vector potential of the incident light field. The absolute value of the vector potential of the pump pulse is 0.32 a.u. which corresponds to an expected electron energy of 1.4 eV. This is in good agreement with the measured electron energies in Fig. S1A. In a next step, we inspect Fig. S1B which shows the same as Fig. S1A but here only the probe pulses are used. The probe pulse's vector potential is 1.37 a.u. which corresponds to an electron energy of 25 eV which is in good agreement with the observed electron kinetic energy in Fig. S1B. This allows for the conclusion that the electron energy can be used to distinguish electrons from the pump and the probe pulse in our pump-probe experiment. Figure S1C shows the same as Fig. S1A and S1B but for the pump-probe experiment. In Fig. S1C an additional peak is seen that is not evident for the pump pulse alone or the probe pulse alone. This peak is at a KER of about 4 eV and an electron energy of about 25 eV. The electron energy of 25 eV indicates that the electron must have been liberated by the probe pulse. The low count rates for a KER of 4 eV in the case of the pump pulse alone and the probe pulse alone show that the peak at a KER of 4 eV is due to molecules that were dissociated to two neutral oxygen atoms with a KER of 4 eV by the pump pulse.

For the sake of completeness, we note that for the cases where the probe pulse is used, we observe a very pronounced peak at a KER of 7.6 eV that is due to the production of two singly charged oxygen atoms (not shown in Fig. S1). In these cases, the two ions are produced at an internuclear distance that is on the order of 1.3 Å, which is the internuclear distance for the molecular ground state (see Fig. 1, A). This leads to significant Coulomb repulsion and the high KER of about 7.6 eV. For the KER peak that is at about 4 eV, the KER of the ions is caused by the pump pulse. The probe pulse liberates the two electrons 1.5 picoseconds later. At this time the internuclear distance is on the order of 75 Å. Thus, the change of the KER that is due to Coulomb repulsion of the ions after ionization is negligible for the KER peak at about 4 eV.

This finding is further illustrated by Fig. S1D. The distributions of the KER from Fig. S1A and S1B are shown as blue and red data points, respectively. The yellow data points are a subset of the data shown in Fig. S1C for which two singly charged oxygen ions were detected. The green data points show another subset from the data shown in Fig. S1C, but here only the cases in which one singly charged oxygen ion was detected is shown. The random coincidences, that belong to the case where two oxygen ions were produced but only one was detected, is subtracted. This subtraction of random coincidences accounts for the finite detection efficiency of our ion detector, which was determined as in Ref. *(32)* and found to be 55%. The experimental data shown in Fig. 3 is processed as the data that belongs to the green and yellow curve in Fig. S1D. For the data shown in Fig. 3 the background from the pump pulse alone and the probe pulse alone was also subtracted after normalization to equal acquisition times. For Fig. 3A (3C) the KER was restricted to 3.6 eV – 4.4 eV (3.7 eV – 4.5 eV). Fig. S3 shows further details regarding the measured electron spectra.

Preparation of the Bell-like state by the pump pulse
The circularly polarized pump pulse at a central wavelength of 390 nm excites the molecule and thereby leads to dissociation via the $1^3\Pi_u$-state as shown in Fig. 1. The expected KER of the two $^3P$ oxygen atoms is 4 eV according to energy conservation for a three-photon absorption. The curves shown in Fig. 1A are taken from Ref. *(19)*. Dissociation to $^1D + {}^3P$ would lead to a kinetic energy release of about 2 eV since the $^1D$-state is 2 eV above the ground state. Thus, we know that the two twin atoms with a total KER of about 4 eV are both in the $^3P$ state.

The ground state of oxygen $X^3\Sigma_g$ possesses an equal number of electrons with $m = -1$ and $m = +1$. Since, dissociation is triggered by a three-photon absorption from the circularly polarized pump pulse, the excitation of the electron must involve two orbitals with different parity. Further, the excitation cannot occur from one $\pi$-orbital to another $\pi$-orbital since the transition occurs from a $\Sigma$-state to a $\Pi$-state. The only remaining possible transitions are the transition from $\pi^*_{2p}$ with $m = +1$ to $\sigma^*_{2p}$ with $m = 0$ and the transition from $\pi^*_{2p}$ with $m = -1$ to $\sigma^*_{2p}$ with $m = 0$. The helicity of the circularly polarized pump pulses defines which transition is favoured (see discussion of $\tilde{q}_{eff}(\alpha)$ below).

Fig. S3 shows that the electron energy spectra contain information on the helicity of the pump pulse and that the quantization axis of the $^3P$ oxygen atoms is defined by the quantization axis of the former molecule until the probe pulse arrives (also see Ref. *(10)*). Depending on the helicity of the circularly polarized pump pulse, the two $^3P$ oxygen atoms in their ground state are described by $\Psi^+_{-00-} = \frac{1}{\sqrt{2}}(|m_{-1}\rangle_A|m_0\rangle_B + |m_0\rangle_A|m_{-1}\rangle_B)$ or by $\Psi^+_{+00+} = \frac{1}{\sqrt{2}}(|m_{+1}\rangle_A|m_0\rangle_B + |m_0\rangle_A|m_{+1}\rangle_B)$.

Expression of the Bell-like state using the quantization axis of the probe pulse
Here we discuss a Bell-like state in the state $\Psi^+_{-00-}$ as in Eq. 1 and note that the conclusions can be drawn for the state $\Psi^+_{+00+}$ in full analogy. The quantization axis of the Bell-like state is given by the quantization axis of the former molecule and the Bell-like state is expressed using the basis $|m_{-1}\rangle, |m_0\rangle, |m_{+1}\rangle$. If this Bell-like state is ionized by a circularly polarized probe pulse at a central wavelength of 780 nm, the light propagation direction of the probe pulse defines the new quantization axis, which is described using the basis $|m'_{-1}\rangle, |m'_0\rangle, |m'_{+1}\rangle$. Let $\alpha$ be the angle between the light's polarization plane and the former molecular axis. Thus, $|90° - \alpha|$ denotes the angle between the two quantization axes (see Fig. 2, A). As shown in Eq. 4, the coefficients a, b, c, d, e, and f can be used to express the prepared Bell-like state in the basis of $|m'_{-1}\rangle, |m'_0\rangle, |m'_{+1}\rangle$. The $\alpha$-dependent, real valued coefficients a, b, c, d, e, and f are defined by Eq. 2 and Eq. 3. Here, $|m_{-1}\rangle, |m_0\rangle, |m_{+1}\rangle$ are the atomic 2p orbitals with $m = -1, m = 0$ and $m = +1$ respectively. Rotation of the states $|m_{-1}\rangle, |m_0\rangle, |m_{+1}\rangle$ by the angle $\alpha$ around an axis that is perpendicular to the quantization axis leads to the new basis $|m'_{-1}\rangle, |m'_0\rangle, |m'_{+1}\rangle$. The values for a, b, c, d, e, and f are shown in Fig. S2A.

The singly ionized oxygen atoms are all in the $^4S^0$ state. The energetically closest, alternatively available final state is the $^2S^0$ state, which is very unlikely to be populated by strong field ionization *(1)* since it has an ionization potential that is 3.3 eV higher than the one of the $^4S^0$ state. Thus, the $^4S^0$ state is the only relevant final state for the singly charged oxygen ions. All 2p electrons in the $^4S^0$ state must have the same spin which justifies the notation in Fig. 1 and Fig. 2. Further, this allows for the conclusion that during the single ionization of a $^3P$ oxygen atom it is only possible to remove the 2p electron that has a spin that is opposite to the spin of all the other 2p electrons as illustrated in Fig. 2.

## Quantitative modeling of the angular dependence of the ionization of only one of the two oxygen atoms

Non-adiabatic tunnel-ionization prefers the liberation of electrons with $|m'_{-1}\rangle$. The probability to liberate an electron with $|m'_{-1}\rangle$ is modeled by $p_-$. For our probe pulse and the ionization potential of atomic oxygen $I_p = 13.62$ eV, the tunneling probability (see Eq. 108 and 109 in Ref. (33)) for an electron with $|m'_{+1}\rangle$ is $p_+ \approx 0.4 \cdot p_-$ and the tunneling probability for an electron with $|m'_0\rangle$ is $p_0 \approx 0.05 \cdot p_-$.

The Bell-like state is described by a two-electron wave function. The probability to find an electron with $m' = -1$ at site A and the other electron with $m' = 0$ at site B is given by $\frac{1}{2}(ae + db)^2$. All possible combinations for detectable combinations follow directly from Eq. 4 and are summarized in Table S1 using the $C$-coefficients. The $C$-coefficients are used to model the occupation of the entangled $m'$-orbitals in the Bell-like state $\Psi^+_{-00-}$ and are expressed via the coefficients a, b, c, d, e, and f. The $C$- coefficients are defined by $C_{xy} \coloneqq |\langle m'_y|_B \langle m'_x|_A \Psi^+_{-00-}\rangle|^2$. The values for the $C$-coefficients as a function of $\alpha$ are shown in Fig. S2B.

The occupation of $|m'_{-1}\rangle$ at site A without any restriction on the wave function at site B is given by $C_{--} + C_{-0} + C_{-+}$. In full analogy the occupation of $|m'_{-1}\rangle$ at site B is $C_{--} + C_{0-} + C_{+-}$. Due to symmetry we know that $C_{--} + C_{-0} + C_{-+} = C_{--} + C_{0-} + C_{+-}$. Accordingly, the occupation of $|m'_0\rangle$ at site A is given by $C_{0-} + C_{00} + C_{0+}$ and that of $|m'_{+1}\rangle$ is $C_{+-} + C_{+0} + C_{++}$. The dependence of the occupation of the $m'$-states at site A as a function of $\alpha$ is shown in Fig. S2C. It should be noted that $C_{--} + C_{-0} + C_{-+} = \frac{1}{2}(ad + da)^2 + \frac{1}{2}(ae + db)^2 + \frac{1}{2}(af + dc)^2 = \frac{1}{2}(a^2 + d^2)$ since $a^2 + b^2 + c^2 = d^2 + e^2 + f^2 = 1$ due to normalization and $a^2d^2 + aedb + afdc = 0$ since $ad + be + cf = 0$ because $|m_{-1}\rangle$ and $|m_0\rangle$ are orthogonal (see Eq. 2 and 3). The fact that $C_{--} + C_{-0} + C_{-+} = \frac{1}{2}(a^2 + d^2)$ is the reason why for single ionization the quantum-mechanical model does not produce deviating predictions compared to the classical model for small values of $p_-$ (see main text).

The probability for single ionization of the Bell-like state $\Psi^+_{-00-}$ at site A and not ionize it at site B is given in Eq. 5 using the definitions $\overline{p_-} = 1 - p_-$, $\overline{p_0} = 1 - p_0$ and $\overline{p_+} = 1 - p_+$.

$$P^{single\_A}_{-00-} = p_-(\overline{p_-}C_{--}+\overline{p_0}C_{-0}+\overline{p_+}C_{-+}) + p_0(\overline{p_-}C_{0-}+ \overline{p_0}C_{00} + \overline{p_+}C_{0+})$$
$$+p_+(\overline{p_-}C_{+-} + \overline{p_0}C_{+0}+\overline{p_+}C_{++}) \qquad (5)$$

Making use of the symmetry with respect to an inversion of the magnetic quantum number, the single ionization of the Bell-like state $\Psi^+_{+00+}$ at site A is modeled by:

$$P^{single\_A}_{+00+} = p_+(\overline{p_+}C_{--} + \overline{p_0}C_{-0} + \overline{p_-}C_{-+}) + p_0(\overline{p_+}C_{0-} + \overline{p_0}C_{00} + \overline{p_-}C_{0+})$$
$$+p_-(\overline{p_+}C_{+-} + \overline{p_0}C_{+0} + \overline{p_-}C_{++}) \qquad (6)$$

The probability to singly ionize the Bell-like state in site A or B, but not ionize it in site A and B, is simply two times the probability to ionize it in site A only and given by $P^{single}_{-00-} = 2 \cdot P^{single\_A}_{-00-}$ and $P^{single}_{+00+} = 2 \cdot P^{single\_A}_{+00+}$.

For $P^{single}_{-00-}$ and $P^{single}_{+00+}$ we have assumed that the pump step (with a given helicity) perfectly populates one of the two possible Bell-like states (as illustrated in Fig. 1). This is of course not the case. Due to symmetry, we know that for $\alpha = 0°$ there must be an equal probability to populate $\Psi^+_{-00-}$ and $\Psi^+_{+00+}$. Let us assume that the probability to prepare the state $\Psi^+_{-00-}$ is given by $q_{eff}(\alpha) = \frac{a^2}{a^2+c^2}$ and the probability to prepare the state $\Psi^+_{+00+}$ is given by $1 - q_{eff}(\alpha) = \frac{c^2}{a^2+c^2}$. The underlying picture for this assumption is that the pump pulse populates a virtual state with $m = -1$ or $m = +1$. The sign of $m$ depends on the helicity of the pump pulse. The virtually excited state is reached absorbing photons from the pump pulse, which has a quantization axis that is given by the pump pulse's propagation direction. However, the molecular axis defines the quantization axis of the dissociating state $1^3\Pi_u$. The coefficients a and c model the projection onto the quantization axis of the molecular state in full analogy to the previous discussion of the coefficients a, b, c, d, e, and f. Using this approach, the trivially expected results, which are $q_{eff}(0°) = 0.5$, $q_{eff}(\alpha) = q_{eff}(-\alpha)$, and $q_{eff}(90°) = q_{eff}(-90°) = 1$ are reproduced, and $q_{eff}$ is monotonic between 0° and 90°. Further, we use the parameter $\eta \in [0,0.5]$ to characterize the $m$-selectivity of the pump step. $\eta=0$ would indicate that molecules that are aligned along the light propagation direction are always dissociating as illustrated in Fig. 1. $\eta=0.5$ would indicate that the pump step produces $\Psi^+_{-00-}$ and $\Psi^+_{+00+}$ with equal probabilities (also see Ref. *(10)*). By using $\tilde{q}_{eff}(\alpha) = 2 \cdot q_{eff}(\alpha) \cdot (0.5 - \eta) + \eta$ instead of $q_{eff}(\alpha)$ we model the angle dependent pump-efficiency in our experiment.

So far, $p_-$ and $\eta$ are the only free parameters of our model. In the following we will introduce two more free parameters. In analogy to Ref. *(34)* we use $\beta$ and additionally $\kappa$ (to take multiphoton absorption into account) to model the $\alpha$-dependent dissociation probability by:

$$D(\alpha) = \left(1 - 0.25 \cdot \beta \cdot (3 \cdot (\cos(\alpha))^2 - 1)\right) \cdot (\cos(\alpha))^\kappa \qquad (7)$$

For the ionization probability of only one of the two oxygen atoms, our quantum-mechanical model leads to the following expressions:

$$P^{single}_{pump\ minus}(\alpha) = D(\alpha) \cdot \left(\tilde{q}_{eff}(\alpha) \cdot P^{single}_{-00-}(\alpha) + \left(1 - \tilde{q}_{eff}(\alpha)\right) \cdot P^{single}_{+00+}(\alpha)\right) \qquad (8)$$

$$P^{single}_{pump\ plus}(\alpha) = D(\alpha) \cdot \left(\tilde{q}_{eff}(\alpha) \cdot P^{single}_{+00+}(\alpha) + \left(1 - \tilde{q}_{eff}(\alpha)\right) \cdot P^{single}_{-00-}(\alpha)\right) \qquad (9)$$

All four free parameters are obtained from comparison of the data for ionization of only one of the two oxygen atoms. We obtain the values $p_- = 0.33, \eta = 0.265, \beta = 1.06$ and $\kappa = 1.74$. The results for $P^{single}_{pump\ minus}$ and $P^{single}_{pump\ plus}$ are plotted as solid lines in Fig. 3B. We optimize $p_-$ and $\eta$ in an outer loop such that the scalar value $\frac{\int P^{single}_{pump\ minus}(\alpha) \cdot \cos(\alpha) d\alpha}{\int P^{single}_{pump\ plus}(\alpha) \cdot \cos(\alpha) d\alpha}$ matches the experimental finding of 1.07 and that $R_{hightheo}=0.5$ (see Ref. *(10)* for details on $R_{hightheo}$). In an inner loop we adjust $\beta$ and $\kappa$ making sure that the mean of $P^{single}_{pump\ minus}(\alpha)$ and $P^{single}_{pump\ plus}(\alpha)$ agrees with the mean of the two experimental curves in Fig. 3A (regardless of an overall normalization factor).

The value for R$_{hightheo}$ is calculated via R$_{hightheo}$=$\frac{\int_{\alpha=42°}^{90°} P_{pump\ minus}^{single}(\alpha)\cdot\cos(\alpha)d\alpha}{\int_{\alpha=42°}^{90°} P_{pump\ plus}^{single}(\alpha)\cdot\cos(\alpha)d\alpha}$ by setting $p_- = 0$, $p_0 = 0$ and $p_+ = 1$. The usage of R$_{hightheo}$ is in full analogy to the procedure used in Ref. *(10)* and builds on the assumption that for very high electron energies only electrons with $m = +1$ contribute the electron energy spectrum. The value of R$_{high}$=0.5 is obtained from the experimental data shown in Fig. S3C and S3F.

Quantitative modeling of the angular dependence of the ionization of both oxygen atoms
In full analogy to the derivation of $P_{-00-}^{single}$ and $P_{+00+}^{single}$, we can express the probability for the double ionization of the two Bell-like states. Double ionization of the Bell-like states means that single ionization occurs at site A and at site B. The double ionization probability for the Bell-like state $\Psi_{-00-}^+$ is described by:

$$P_{-00-}^{double} = p_-(p_- C_{--} + p_0 C_{-0} + p_+ C_{-+})$$
$$+ p_0(p_- C_{0-} + p_0 C_{00} + p_+ C_{0+}) + p_+(p_- C_{+-} + p_0 C_{+0} + p_+ C_{++}) \quad (10)$$

Making use of the symmetry with respect to the inversion of the sign of the magnetic quantum number, the double ionization probability of the Bell-like state $\Psi_{+00+}^+$ is modeled by:

$$P_{+00+}^{double} = p_+(p_+ C_{--} + p_0 C_{-0} + p_- C_{-+})$$
$$+ p_0(p_+ C_{0-} + p_0 C_{00} + p_- C_{0+}) + p_-(p_+ C_{+-} + p_0 C_{+0} + p_- C_{++}) \quad (11)$$

In full analogy to $P_{pump\ minus}^{single}$ and $P_{pump\ plus}^{single}$, we define:

$$P_{pump\ minus}^{double} = D(\alpha) \cdot \left(\tilde{q}_{eff}(\alpha) \cdot P_{-00-}^{double}(\alpha) + \left(1 - \tilde{q}_{eff}(\alpha)\right) \cdot P_{+00+}^{double}(\alpha)\right) \quad (12)$$

$$P_{pump\ plus}^{double} = D(\alpha) \cdot \left(\tilde{q}_{eff}(\alpha) \cdot P_{+00+}^{double}(\alpha) + \left(1 - \tilde{q}_{eff}(\alpha)\right) \cdot P_{-00-}^{double}(\alpha)\right) \quad (13)$$

The results from Eq. 12 and 13 are shown as solid lines in Fig. 3D and show very good agreement with the experimental data presented in Fig. 3C. It is important to note that our model has four free parameters ($p_- = 0.33, \eta = 0.265, \beta = 1.06$ and $\kappa = 1.74$). All four free parameters are determined using the experimental results for the single ionization of the Bell-like state. Further, we note that the contributions from $\Psi_{-00-}^+$ and $\Psi_{+00+}^+$ to the total ionization probability are added up incoherently within our model (in full analogy to Ref. *(10)*). We expect that interferences that are due to a coherent summation of both channels might affect the electron angular distribution in the molecular frame but would not produce different results for the total yield, which is the integral over all possible electron emission directions. The good predictive power regarding the double ionization of the Bell-like state supports the validity of the quantum-mechanical model (compare Fig. 3C with solid lines in Fig. 3D).

Quantitative modeling using a classical model
For comparison with our quantum-mechanical model we use a model, which we refer to as classical model. In contrast to the quantum-mechanical model, which uses a Bell-like state (see Eq. 1), the classical model uses a classical state which is in 50% of the cases described by $|m_{-1}\rangle_A|m_0\rangle_B$ and in the other 50% it is modeled by $|m_0\rangle_A|m_{-1}\rangle_B$. (The classical model for $\Psi_{+00+}^+$ is defined accordingly.) Besides this difference regarding the definition of the initial state, the classical model is the same as the quantum-mechanical model. In full analogy to the procedure for the quantum-mechanical model, this leads to different $C$-coefficients using the definition $C_{xy}^{classical} := \frac{1}{2}|\langle m'_y|_B \langle m'_x|_A |m_{-1}\rangle_A |m_0\rangle_B |^2 + \frac{1}{2}|\langle m'_y|_B \langle m'_x|_A |m_0\rangle_A |m_{-1}\rangle_B |^2$. The coefficients are $C_{--}^{classical} = \frac{1}{2}(ad)^2 + \frac{1}{2}(da)^2$, $C_{0-}^{classical} = \frac{1}{2}(bd)^2 + \frac{1}{2}(ea)^2$, $C_{+-}^{classical} = \frac{1}{2}(cd)^2 + \frac{1}{2}(fa)^2$, $C_{-0}^{classical} = \frac{1}{2}(ae)^2 + \frac{1}{2}(db)^2$, $C_{00}^{classical} = \frac{1}{2}(be)^2 + \frac{1}{2}(eb)^2$, $C_{+0}^{classical} = \frac{1}{2}(ce)^2 + \frac{1}{2}(fb)^2$, $C_{-+}^{classical} = \frac{1}{2}(af)^2 + \frac{1}{2}(dc)^2$, $C_{0+}^{classical} = \frac{1}{2}(bf)^2 + \frac{1}{2}(ec)^2$, and $C_{++}^{classical} = \frac{1}{2}(cf)^2 + \frac{1}{2}(fc)^2$. These classical $C$-coefficients are used in full analogy to the quantum-mechanical $C$-coefficients to produce the results that are shown as dashed lines in Fig. 3B and 3D. The usage of a classical state instead of $\Psi_{-00-}^+$ or $\Psi_{+00+}^+$ excludes entanglement and limits the correlations to classical correlations. The sum over all nine $C^{classical}$-coefficients is 1 for all values of $\alpha$, which ensures that also in the classical model the normalization does not depend on $\alpha$. The parameters for the classical model are determined independently but in full analogy to the quantum-mechanical model and found to be $p_- = 0.61, \eta = 0.285, \beta = 1.04$ and $\kappa = 1.84$. (As for the quantum-mechanical model, also here, the four parameters are found using only the data from the events that are shown in Fig. 3, A.) The result for single ionization of only one of the atoms is shown as dotted lines in Fig. 3B and shows only minute differences compared to the quantum-mechanical prediction. It should be noted that the minute differences of the quantum-mechanical model and the classical model for single ionization that are seen in Fig. 3B would vanish if $p_-$ approached zero. However, also in this case the differences for double ionization (see Fig. 3D) would remain.

The classical model for double ionization is realized as described above by using the $C^{classical}$-coefficients instead of the $C$-coefficients. The results from the classical model are shown as dotted lines in Fig. 3D and show inferior agreement with the experimental results (Fig. 3C).

The phase $\phi$ of the Bell-like state
In principle, there can be a relative phase $\phi$ between the two parts of the electronic wave function of the Bell-like states $\Psi_{-00-}^+$ and $\Psi_{+00+}^+$. For $\Psi_{-00-}^+$ this results in the expression $\frac{1}{\sqrt{2}}(|m_{-1}\rangle_A |m_0\rangle_B + e^{i\phi}|m_0\rangle_A |m_{-1}\rangle_B)$. In our experiment, the phase $\phi$ must be close to zero which leads to $e^{i\phi} = 1$. This is evident from comparing the experimental result with the predictions from the quantum-mechanical model for different values of $\phi$. As a reference we use the result from our quantum-mechanical model for which we have used $\phi = 0°$ (see Fig. 3A and 3C). For comparison we show the corresponding results for $\phi = 45°$, $\phi = 90°$ and $\phi = 110°$ in Fig. S4. To this end the quantum-mechanical model is employed and the definition of the $C$-coefficients is generalized using the expression $C_{xy}^{phase} := \frac{1}{2}|\langle m'_y|_B \langle m'_x|_A |m_{-1}\rangle_A |m_0\rangle_B + e^{i\phi}\langle m'_y|_B \langle m'_x|_A |m_0\rangle_A |m_{-1}\rangle_B |^2$, which is equivalent to multiplying the second summand in Table S1 with $e^{i\phi}$. For each value of $\phi$ the data for single ionization (Fig. 3A) is used to find the parameters $p_-, \eta, \beta$ and $\kappa$ (see caption of Fig. S4 for the corresponding values). Interestingly, for $\phi = 90°$ the quantum-mechanical model yields the same result as the classical model (compare dashed lines in Fig. 3B and 3D with Fig. S4B and S4E). It is found that for $|\phi| < 90°$ ($|\phi| > 90°$) the double-hump structure for double ionization is more (less) pronounced compared to the classical model. Comparison of Fig. S4 with the experimental result in Fig. 3 shows that $\phi$ must be close to zero. This leads to a symmetric spatial wave function of the prepared Bell-like states which implies an antisymmetric spin wave function (in agreement with the illustrations in Fig. 1 and 2).

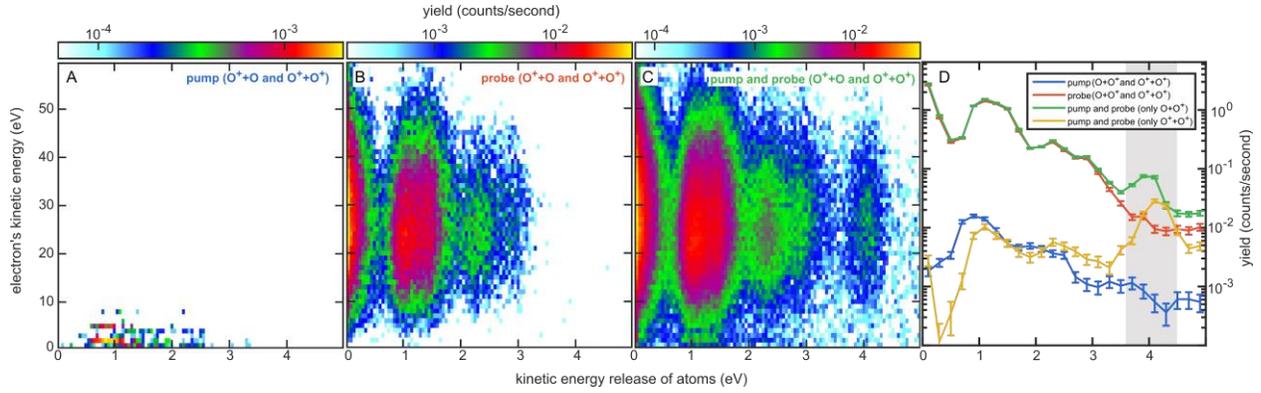

**Fig. S1. (A)** The experimentally measured rate to detect at least one oxygen ion is shown as a function of the kinetic energy of the detected electron and the kinetic energy release (KER) of the atoms for the case in which only the pump pulse is applied. **(B)** The same as A but here only the probe pulse is applied. **(C)** The same as A and B but here the pump and the probe pulse are both applied. **(D)** The blue and red line show the same data as in A and B. The yellow line shows the experimentally determined rate for the ionization of both oxygen atoms by the probe pulse. The green line shows the experimentally determined rate for the ionization of one of the two oxygen atoms by the probe pulse after subtracting random coincidences, that are due to cases in which both oxygen atoms are ionized but only one is detected. All data shown here were measured using a pump pulse with an anticlockwise rotating laser electric field (corresponds to the preparation in the $\Psi^+_{-00-}$ -state). The gray shaded area in D highlights the energy range around a KER of 4 eV. It should be noted that the events with a KER that is smaller than 3.5 eV mainly belong to the production of $O^+ + O$ and the events with a KER that is higher than 3.5 eV belong to the production of $O^+ + O^+$. The channel $O^+ + O^+$ in B and C has another pronounced maximum at a KER of 7.6 eV (not shown, see text for details). See Fig. S3 for details regarding the electron energy spectra.

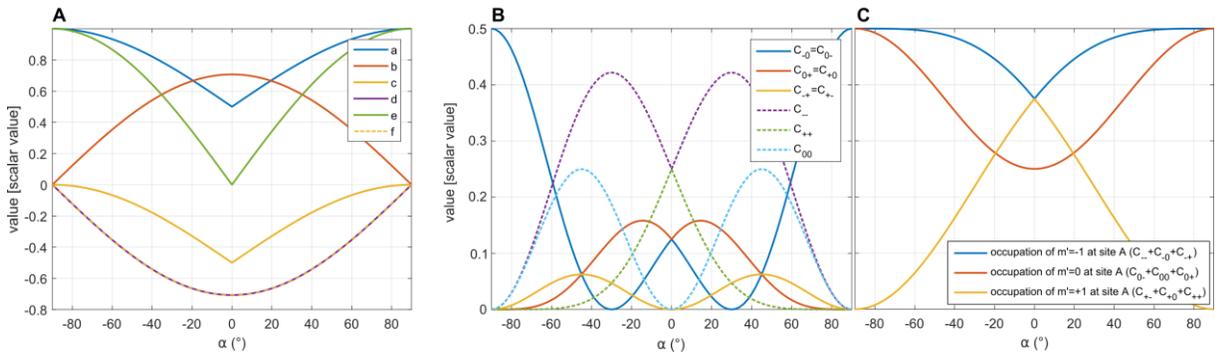

**Fig. S2. (A)** Values of the coefficients a, b, c, d, e, and f, that are used to express the Bell-like state using the quantization axis of the probe pulse, are shown as a function of $\alpha$. **(B)** Values of the $C$-coefficients that represent the occupation of the different entangled $m'$-orbitals. The sum over the nine $C$-coefficients is 1 for all values of $\alpha$. **(C)** Occupation of the different uncorrelated $m'$-orbitals. $\alpha$ is the intermediate angle of the former molecular axis and the light's polarization plane.

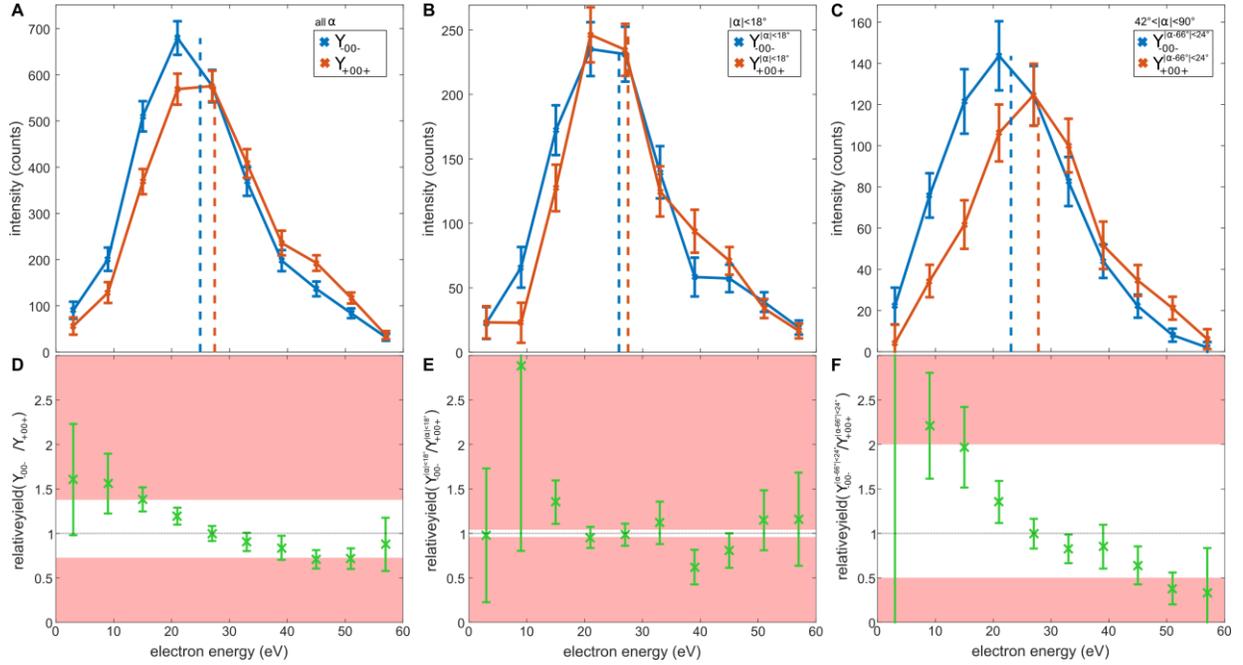

**Fig. S3.** **(A)** $Y_{-00-}$ is the experimentally obtained electron energy spectrum that is measured in coincidence with the data shown in Fig. 3A that is labeled with $\Psi^+_{-00-}$. $Y_{+00+}$ is the same for the data labeled with $\Psi^+_{+00+}$. **(B)** [**(C)**] The same as A but for the subset of events with $|\alpha| < 18°$ [$|\alpha| > 42°$]. The vertical dashed lines in A-C indicate the corresponding mean values of the measured electron energy distributions. **(D-F)** The ratios $Y_{-00-}/Y_{+00+}$ from the data that is shown in A-C. It is seen in F that the ratio reaches values of down to $R_{high}=0.5$ for high electron energies. The red shaded areas in D-F indicate ratios that are not expected within the quantum-mechanical model for the parameters that are used ($p_- = 0.33, \eta = 0.265, \beta = 1.06$ and $\kappa = 1.74$).

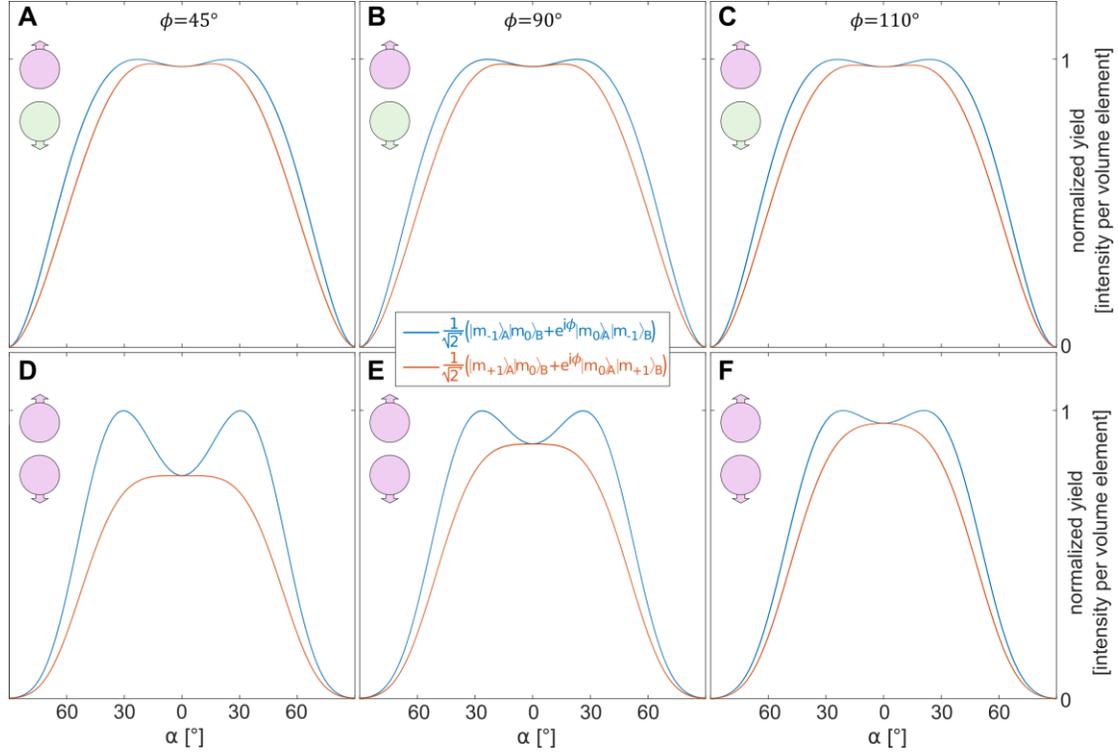

**Fig. S4. (A)** and **(D)** The results from the quantum-mechanical model (in analogy to the solid lines in Fig. 3B and 3C) but here an initial state $\frac{1}{\sqrt{2}}(|m_{-1}\rangle_A|m_0\rangle_B + e^{i\phi}|m_0\rangle_A|m_{-1}\rangle_B)$ is used for the blue line and an initial state $\frac{1}{\sqrt{2}}(|m_{+1}\rangle_A|m_0\rangle_B + e^{i\phi}|m_0\rangle_A|m_{+1}\rangle_B)$ is used for the red line with a phase of $\phi = 45°$. For this scenario the parameters $p_- = 0.37, \eta = 0.268, \beta = 1.05$ and $\kappa = 1.75$ are found using the same procedure as for Fig. 3. **(B)** and **(E)** The same as A and D but a phase of $\phi = 90°$ is used and the parameters $p_- = 0.61, \eta = 0.285, \beta = 1.04$ and $\kappa = 1.84$ are found. The result is equivalent to the result for the classical model that is shown as dashed lines in Fig. 3B and 3D. **(C)** and **(F)** The same as A and D but a phase of $\phi = 110°$ is used and the parameters $p_- = 0.84, \eta = 0.303, \beta = 1.01$ and $\kappa = 1.92$ are found. A-D show results for the single ionization and D-F show results for the double ionization of the Bell-like state.

|  | $|m'_{-1}\rangle_A$ | $|m'_0\rangle_A$ | $|m'_{+1}\rangle_A$ |
|---|---|---|---|
| $|m'_{-1}\rangle_B$ | $C_{--} = \frac{1}{2}|ad + da|^2$ | $C_{0-} = \frac{1}{2}|bd + ea|^2$ | $C_{+-} = \frac{1}{2}|cd + fa|^2$ |
| $|m'_0\rangle_B$ | $C_{-0} = \frac{1}{2}|ae + db|^2$ | $C_{00} = \frac{1}{2}|be + eb|^2$ | $C_{+0} = \frac{1}{2}|ce + fb|^2$ |
| $|m'_{+1}\rangle_B$ | $C_{-+} = \frac{1}{2}|af + dc|^2$ | $C_{0+} = \frac{1}{2}|bf + ec|^2$ | $C_{++} = \frac{1}{2}|cf + fc|^2$ |

**Table. S1.** Overview of the $C$-coefficients (based on Eq. 4).